# PXIE OPTICS AND LAYOUT[*]

V. Lebedev[#], S. Nagaitsev, J.-F. Ostiguy, A. Shemyakin, N. Solyak, B. Shteynas
Fermilab, Batavia, IL 60510, USA

*Abstract*

The Project X Injector Experiment (PXIE) will serve as a prototype for the Project X front end. The aim is to validate the Project-X design [1] and to decrease technical risks mainly related to the front end. The paper discusses the main requirements and constraints motivating the facility layout and optics. Final adjustments to the Project X front end design, if needed, will be based on operational experience gained with PXIE.

## MACHINE LAYOUT

PXIE [2] will accelerate H$^-$ 1 mA (average) beam to about 25 MeV. Figure 1 presents the PXIE layout. The total length of the accelerator is about 40 m. It will consist of a 30 kV 5 mA H$^-$ DC ion source, a low energy beam transport section (LEBT), 2.1 MeV 162.5 MHz CW RFQ, a medium energy beam transport section (MEBT), two SC cryomodules, a diagnostic section and a beam dump. The first cryomodule has 8 half-wave (HW) SC cavities operating at 162.5 MHz frequency with 8 focusing SC solenoids located between cavities. The second cryomodule has 8 single spoke cavities (SSR1) operating at 325 MHz with 4 SC focusing solenoids. An RF separator is located at the beginning of diagnostics section (after the second cryomodule) to support studies of beam extinction for removed bunches. Its frequency of 243.75 MHz is chosen to be 1.5 times of bunch frequency so that "even" and "odd" bunches are deflected in opposite directions. The dipole of the magnetic spectrometer bends the beam down by $20^0$. A 50 kW beam dump is located after the dipole. Detailed description of PXIE systems can be found in Ref. [2].

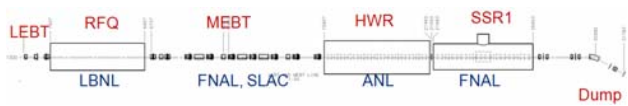

Figure 1: PXIE layout.

## PXIE OPTICS

Measurements of ion source beam emittance resulted in the beam normalized rms emittance to be within 0.09-0.12 mm mrad for a current range 2 to 10 mA [3]. The LEBT section transports the beam to the RFQ. It includes a switching dipole and normal conduction solenoids to match the beam phase space to the RFQ. The edge focusing of switching dipole is adjusted to minimize asymmetry between its horizontal and vertical focusing. The beam space charge introduces non-linear focusing which strongly effects single particle motion and results in emittance growth. To mitigate it space charge compensation by residual gas ions will be used.

Bunching and acceleration in the RFQ was designed to minimize the beam loss and the emittance growth in the current range of 4 to 10 mA [4]. Simulations in the current range 4 to 10 mA yield the normalized rms emittances at the RFQ exit to be 0.15 mm mrad and 0.22-0.25 mm mrad for transverse and longitudinal planes, correspondingly. RFQ vanes are extended by a few centimetres to make a matcher reducing transverse β-functions in the MEBT matching section.

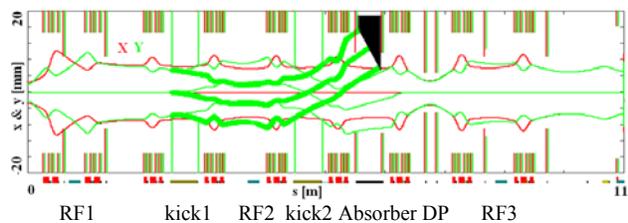

Figure 2: 3σ beam envelopes ($\varepsilon_{rms\_n}$=0.25 mm mrad) for accepted and removed bunches through MEBT; red – horizontal plane, green – vertical plane, red and green vertical lines show aperture limitations for *x* and *y* planes.

Figure 2 presents beam envelopes through the MEBT. Triplet focusing with ~90 deg. phase advance per cell was chosen for the MEBT. It minimizes the beam size and creates sufficiently "smooth" focusing resulting in sufficiently small emittance growth. To obtain acceptable voltage (power) for the bunch-by-bunch MEBT kickers they are separated by 180 deg in betatron phase. It minimizes the gap between kicker plates and the required driving voltage. Each kicker [5] has 16 mm gap with 13 mm aperture restriction, 500 mm length, and is driven differentially by two power amplifiers with ±250 V voltage each. Bunches deflected down (see Figure 2) pass the beam absorber and proceed for further acceleration. Bunches deflected up (shown by thick green line in Figure 2) are stopped at the beam absorber. The drift section immediately following the absorber section has a reduced aperture (10 mm) to introduce effective differential pumping required to prevent performance degradation of the SC cavities due to high gas load from the absorber. The MEBT has three 162.5 MHz normal conducting cavities to prevent beam debunching and to match the longitudinal beam envelope between RFQ and SC linac. The maximum accelerating RF voltage is 100 kV (amplitude).

To fully exploit the high value of RF voltage delivered by SC cavities the accelerating structure has to be compact. This requirement determines the optics structure for the cryomodules. Both of them have solenoidal focusing which in the 2 to 30 MeV energy range makes more compact focusing than quadrupoles. High

---

accelerating gradient in the first cryomodule where energy gain per cavity is comparable with the energy itself creates two problems. The first one is related to longitudinal overfocusing and the second one to transverse defocusing by e.-m. fields of the cavities. The transverse defocusing depends on a particle position in a bunch and therefore it requires stronger focusing than one would need for reference particle focusing (see Figure 3). These considerations led to the choice that an optical cell of the first cryomodule includes one cavity per solenoid. In spite of that compact focusing structure the accelerating voltage for the first three cavities of the HW cryomodule is reduced (see Figure 4) so that the longitudinal phase advance would not significantly exceed 90 degree.

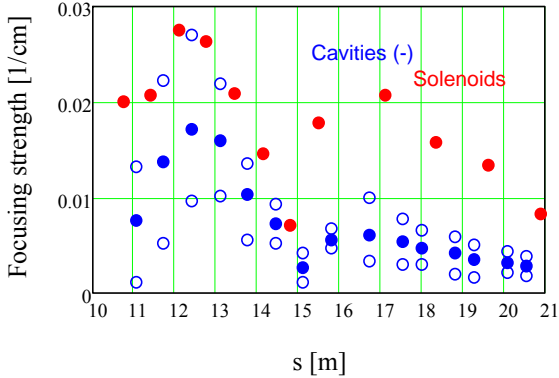

Figure 3: Focusing strength of solenoids (red) and nearby SC cavities for the reference particle (solid blue) and ±4σ longitudinal bunch ends. Sign of the cavity focusing is changed from negative to positive.

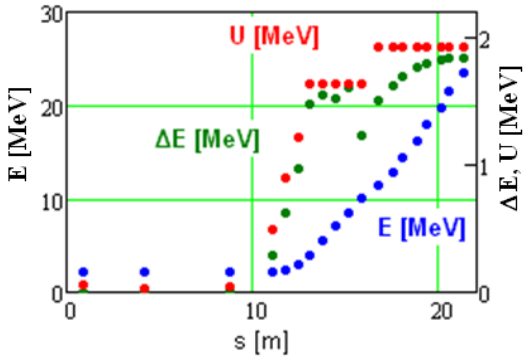

Figure 4: Dependences of beam energy (blue), cavity voltage (red) and actual beam energy change in a cavity (green) on cavity longitudinal coordinates.

Both the longitudinal overfocusing and the transverse defocusing decrease with beam energy. That allows reduction in the number of focusing solenoids in the second cryomodule where one solenoid follows after two cavities. The relative strengths of space charge effects for both transverse and longitudinal planes are about the same through the beam acceleration. However longitudinal dynamics is additionally affected by the strong non-linearity of the accelerating field focusing; therefore a diligent approach is required for the treatment of longitudinal motion. To reduce the longitudinal motion perturbation at the transition between cryomodules the HW cryomodule is ended with a cavity and the SSR1 cryomodule starts from a cavity. Thus the cryomodules have the following structure: (S-C-S-C-S-C-S-C-S-C-S-C-S-C-S-C) for the HW and (C-S-CC-S-CC-S-CC-S-C) for the SSR1, where C stands for a cavity and S for a solenoid. Using a solenoid as the first element in the HW cryomodule also improves differential pumping between the beam absorber and SC cavities due to cryo-pumping of a cold vacuum chamber located in the solenoid. Its low temperature (2K°) results in good cryo-pumping for all gases including hydrogen. Figure 5 presents the bunch envelopes through PXIE (from the end of RFQ to the beam dump). Figure 6 shows corresponding accelerating phases and the bunch length inside RF cavities (expressed in degrees of corresponding cavity phase).

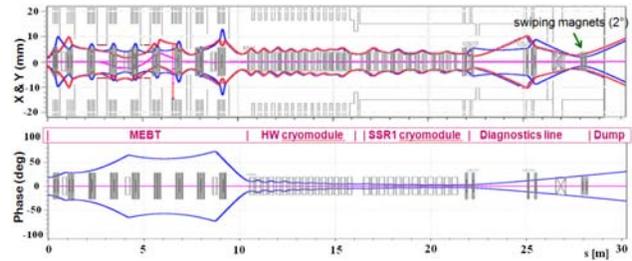

Figure 5: 3σ bunch sizes from RFQ end to beam dump; top - blue and red lines represent horizontal and vertical planes, correspondingly; bottom - 3σ bunch length.

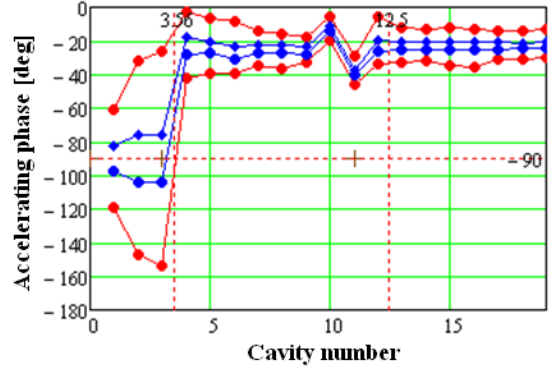

Figure 6: Accelerating phases for 1σ and 4σ bunch boundaries; vertical lines mark cryomodule boundaries.

Figure 7 presents results of simulations of the beam emittance evolution. An emittance growth for both transverse planes is moderate and is within specifications. There is significantly larger growth for the longitudinal emittance mainly related to longitudinal focusing non-linearity. The requirements on the longitudinal emittance growth in Project X are not too strict. However if necessary the emittance growth can be reduced by lowering an accelerating rate. Note also that non-linearities introduced by different cavities can be compensated by minor adjustments of accelerating phases. That presents considerable freedom for further possible improvements.

We also studied the implication of a single cavity failure on machine operation. Calculations show that machine operation remains possible with acceptable performance degradation even when the missing cavity is the lowest energy one.

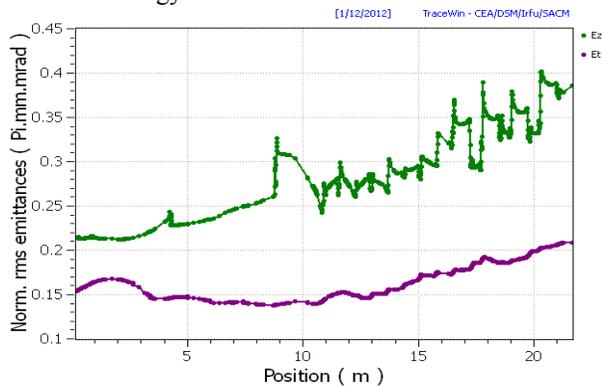

Figure 7: Dependence of longitudinal (green) and transverse (red and blue) rms emittances on longitudinal coordinate from the RFQ exit to the SSR1 end for 5 mA beam.

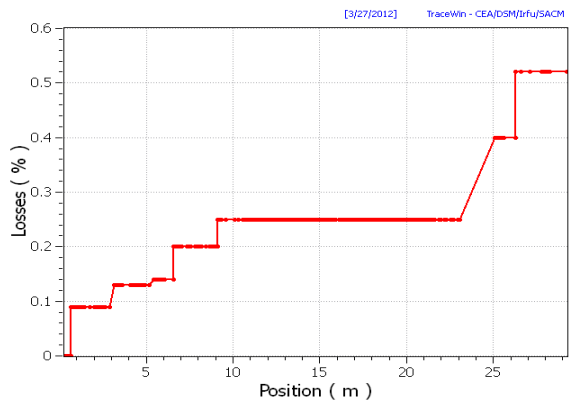

Figure 8: Integrated particle loss anong PXIE.

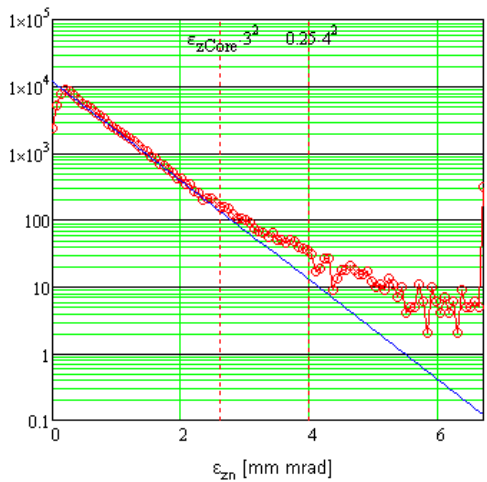

Figure 9: Particle longitudinal distribution at the end of RFQ simulated for 5 mA beam current.

## BEAM LOSS AND BEAM EXTINCTION

Figure 8 shows the integrated particle loss along the machine. There are two areas where particle loss is observed: the MEBT ($s \le 10$ m) and the diagnostic section ($s \ge 22$ m). In the both areas the loss happens at the designated transverse scrapers. The particle loss from the accelerating bucket is more than an order of magnitude smaller. It happens due to long non-Gaussian tails of the RFQ longitudinal distribution (see Figure 9). In contrast to the transverse tails it is close to impossible to intercept the major fraction of longitudinal tails at designated scrapers inside the SC cryomodules. Simulations show that particles in far tails are lost extremely fast. If a particle slips out of acceleration its energy deviates significantly after passing just one cavity. Then, overfocusing in the downstream solenoid results in its loss in the next cavity. Simulations show that even if beam collimators would be installed near each solenoid they cannot intercept major fraction of longitudinal beam loss. Fortunately the RFQ tails are expected to be sufficiently small and should not yield the beam loss exceeding few watts at cryogenic surfaces.

Some of Project X experiments may require extremely good extinction for removed bunches. The target value is smaller than $10^{-9}$, *i.e.* much less than one particle per bunch. A finite population of longitudinal RFQ tails can be the main limitation for achieved beam extinction. Although there is very good rejection of the tails in the first SC cryomodule the weak longitudinal focusing in MEBT and its large length allow momentum tails of allowed bunches to move to the nearby rejected bunches and be accepted in their RF bucket. Figure 10 presents the phase space at the bunch chopper for particles which will be accepted for further acceleration. Lanes above and below the main bunch will be accepted in the nearby buckets. Thus, the longitudinal RFQ tails suppression is the key to achieve good extinction for removed bunches.

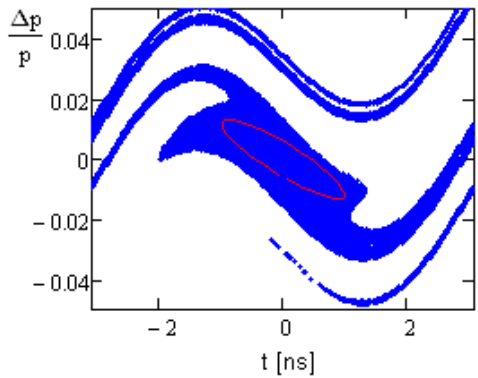

Figure 10: Longitudinal phase space at the beam chopper location for particles accepted for further acceleration (blue dots). Red line shows 4σ bunch boundary.

## REFERENCES

[1] Project X Reference Design Report, http://projectx-docdb.fnal.gov/cgi-bin/ShowDocument?docid=776.
[2] S. Nagaitsev et al., "PXIE: Project X injector Experiments", these proceedings.
[3] Q. Ji, "PXIE LEBT Concept", https://indico.fnal.gov/conferenceOtherViews.py?view=standard&confId=5300, (2012).
[4] Steve Virostek et al., "Design and Analysis of the PXIE CW Radio-frequency Quadrupole", these proceedings.
[5] V. Lebedev et al. "Progress with PXIE MEBT Chopper", these proceedings.